\documentstyle[12pt]{article}
\def\da{{\dot \alpha}} \def\dbe{{\dot \beta}}

\def\spin#1#2{\omega_{#2}{}^{#1}}
\def\RS #1#2{\psi_{#2}{}^{#1}}
\def\bRS #1#2{\5\psi_{#2}{}^{#1}}
\def\viel#1#2{e_{#2}{}^{#1}}
\def\Viel#1#2{E_{#1}{}^{#2}}

\def\s0#1#2{\mbox{\small{$\frac{#1}{#2}$}}}
\def\5{\bar }  \def\6{\partial } \def\7{\hat } \def\4{\tilde }

\def\bea{\begin{eqnarray}} \def\eea{\end{eqnarray}}
\def\beann{\begin{eqnarray*}} \def\eeann{\end{eqnarray*}}
\def\beq{\begin{equation}} \def\eeq{\end{equation}}
\def\ba{\begin{array}} \def\ea{\end{array}}
\def\ben{\begin{enumerate}} \def\een{\end{enumerate}}

\newcommand{\mysection}[1]{\section{#1}
            \setcounter{equation}{0}\setcounter{figure}{0}}

\begin{document}
\begin{flushright}
UB--ECM--PF 97/01\\
hep-th/9704046
\end{flushright}

\begin{center}
{\Large {\bf Deformed supergravity with local $R$-symmetry}}
\end{center}

 \begin{center}
 {\large
 Friedemann Brandt}
 \end{center}

 \begin{center}{\sl
 Departament d'Estructura i Constituents de la Mat\`eria,
 Facultat de F\'{\i}sica,
 Universitat de Barcelona,
 Diagonal 647,
 E-08028 Barcelona, Spain.\\
 E-mail: brandt@ecm.ub.es
 }\end{center}

\begin{abstract}
Using deformation theory based on BRST cohomology, a supergravity model is 
constructed which interpolates through a continuous deformation parameter 
between new minimal supergravity with an extra $U(1)$ gauge multiplet and 
standard supergravity with local $R$-symmetry in a formulation with a 
nonstandard set of auxiliary fields. The deformation implements an 
electromagnetic duality relating the extra $U(1)$ to the $R$-symmetry.
A consistent representative of the $R$-anomaly in the model is proposed too.
\end{abstract}

\mysection{Introduction}\label{intro}

This paper reports a result that arose from a BRST-cohomological
analysis \cite{sugra} of so-called ``old minimal'' and 
``new minimal'' supergravity (SUGRA) \cite{old,new}, including their
coupling to Yang--Mills gauge multiplets. One of the questions
addressed in \cite{sugra} was the classification of the
possible nontrivial consistent deformations of these models.
Such deformations may change simultaneously
the Lagrangian and the form and algebra of
the gauge transformations in a continuous manner,
such that the deformed action is invariant under the deformed gauge 
transformations.
A deformation is called trivial if it represents merely
a local field redefinition. 

Of course, both old and new minimal SUGRA, like most gauge theories, 
have infinitely many nontrivial deformations. However, 
most of these deformations change only the action but
not the gauge transformations \cite{sugra}. Such deformations
thus add further terms to the action
which are invariant under the original gauge transformations, such
as invariants involving higher powers in the curvatures 
and their derivatives. They are of interest, for instance, within a
conventional perturbative quantization approach, for they
provide invariant candidate counterterms.
The classification and construction of all these terms is
described in \cite{sugra}.

On the other hand, deformations which
do change nontrivially the gauge transformations are rather
exceptional. For instance,
they do not exist at all in old minimal SUGRA coupled (only) to
Yang--Mills gauge multiplets whenever the
corresponding gauge group is semisimple \cite{sugra}. 
Of course it is worthwhile to look for
deformations which change the gauge transformations 
in a nontrivial way, for they might provide novel classical 
SUGRA models, or even occur as quantum deformations 
of known classical models.

One of these exceptional deformations is presented in this paper.
It was announced already in \cite{sugra} and has rather
unusual features. Namely, it
deforms new minimal SUGRA into standard (``old'') minimal SUGRA with
local $R$-symmetry. More precisely, it involves
a deformation parameter, denoted by $g_1$ throughout the paper,
such that $g_1=0$ reproduces new minimal SUGRA with an extra
local $U(1)$ symmetry (different from the $R$-symmetry),
whereas for all nonvanishing values of $g_1$
the deformed model is equivalent to
old minimal SUGRA with local $R$-symmetry, at least classically. 
The latter equivalence holds on-shell, after suitable local field 
redefinitions which make sense only for $g_1\neq 0$ 
and provide a new set of auxiliary fields closing 
the supersymmetry (SUSY) algebra off-shell.

The resulting model thus has the remarkable property to
incorporate two different SUGRA models and to connect them
{\em continuously} by means of a 
coupling constant (deformation parameter).
The deformation exists thanks to the presence of a
2-form gauge potential in new minimal SUGRA and 
requires the coupling of new minimal SUGRA 
to an extra $U(1)$ gauge multiplet.
The extra $U(1)$ symmetry disappears effectively 
for $g_1\neq 0$ via an electromagnetic duality
which relates it to the $R$-symmetry and 
is implemented by the deformation itself.
It might be instructive to examine whether
this result can be related to the standard
constructions \cite{dual1,dual2} relating
old and new minimal SUGRA by seeming different
duality transformations.

The paper has been organized as follows.
Section \ref{sketch}
first reviews briefly the construction \cite{bh} of consistent
deformations by BRST cohomological means in general, 
and then illustrates the specific computation performed here
for a very simple toy model discussed already in \cite{sugra}.
The description of the technically more
involved calculation in SUGRA is
relegated to the appendix. Section \ref{new} presents
the resulting deformation of new minimal SUGRA
coupled only to the extra $U(1)$ gauge multiplet.
The relation of this model to standard (old minimal)
SUGRA with local $R$-symmetry is established in section \ref{disc}
which also contains a discussion of the above-mentioned 
electromagnetic duality. In section \ref{gen} extensions of
the model are discussed, 
namely the inclusion of a Fayet--Iliopoulos term \cite{FI} for 
the extra $U(1)$ symmetry and the coupling to matter and further 
Yang--Mills gauge multiplets. Finally, in section \ref{anos},
the question of anomalies associated with the $R$-symmetry 
of our model is addressed and
a consistent (= BRST invariant) representative of ``the''
$R$-anomaly is proposed in explicit form. 
I use the same conventions as in \cite{sugra}.

\mysection{Sketch of the computation}\label{sketch}

\subsection{Consistent deformations and BRST cohomology}\label{method}

The relation of consistent deformations of gauge theories to
the BRST cohomology was pointed out by Barnich and
Henneaux in \cite{bh}. It is based on the field-antifield formalism
of Batalin and Vilkovisky \cite{bv}. The idea is to deform
the solution of the (classical) master equation which is 
the central quantity of this formalism. To that end one departs
from the solution ${\cal S}^{(0)}$ of the master equation in
the original (undeformed) theory and looks 
for a deformed solution ${\cal S}$ of the form
\beq {\cal S}={\cal S}^{(0)}+g\, {\cal S}^{(1)}
              +\s0 12\, g^2{\cal S}^{(2)}+\dots
\label{me1}\eeq
where $g$ is a deformation parameter which plays the
role of a coupling constant in the deformed theory.
The master equation for ${\cal S}$,
\beq
\left({\cal S},{\cal S}\right)=0,
\label{me2}\eeq
is then decomposed into parts with definite degrees in $g$, leading
to a tower of equations which are investigated one after another,
\beq
\left({\cal S}^{(0)},{\cal S}^{(0)}\right)=0,\quad 
\left({\cal S}^{(0)},{\cal S}^{(1)}\right)=0,\quad 
\left({\cal S}^{(1)},{\cal S}^{(1)}\right)
+\left({\cal S}^{(0)},{\cal S}^{(2)}\right)=0,
\quad\cdots\ .
\label{me3}\eeq
The first of these equations is the master equation for the 
original (undeformed) model. The second one, together
with the nontriviality of the sought deformations, 
requires ${\cal S}^{(1)}$ to 
represent a nontrivial cohomology class of the local BRST cohomology
in the original theory. More precisely, 
${\cal S}^{(1)}$ is determined by
$H^0(s^{(0)})$, the cohomology of the original
(undeformed) BRST operator $s^{(0)}$ on local
functionals of the fields and antifields at ghost number zero%
\footnote{$s^{(0)}$ is strictly nilpotent on all 
fields and antifields and  generated in the antibracket by 
${\cal S}^{(0)}$ according to 
$s^{(0)}\ \cdot\ =({\cal S}^{(0)},\ \cdot\ )$.
Analogously the BRST operator in the deformed theory 
is denoted by $s$ and generated by ${\cal S}$ according to
$s\ \cdot\ =({\cal S},\ \cdot\ )$.}. 
The subsequent equations in (\ref{me3}) can
further obstruct the construction of ${\cal S}$ through
$H^1(s^{(0)})$, the cohomology of $s^{(0)}$ at ghost number one.
For instance, the third equation (\ref{me3}) requires 
$({\cal S}^{(1)},{\cal S}^{(1)})$
to be trivial (BRST-exact) in $H^1(s^{(0)})$
(it is $s^{(0)}$-closed by the second equation (\ref{me3}),
thanks to the Jacobi identity for the antibracket).

A study of the BRST cohomology at ghost numbers zero and one therefore
allows  to classify and construct systematically the various  
consistent deformations of a given gauge theory.
(\ref{me1}) is of course only a special case of 
deformations of the form
\beq {\cal S}={\cal S}^{(0)}+\sum_ig_i{\cal S}^{(1)}_i
+\frac 12\sum_{ij}g_ig_j{\cal S}^{(2)}_{ij}+\dots
\label{me4}\eeq
which can be analysed analogously to
(\ref{me3}). In particular, the ${\cal S}^{(1)}_i$ 
represent inequivalent cohomology classes 
of $H^0(s^{(0)})$.

In our case, ${\cal S}^{(0)}$ is of the form
\beq {\cal S}^{(0)}=\int d^4x\, 
\left({\cal L}^{(0)}-(s^{(0)}\Phi^A)\Phi^*_A\right)
\label{me5}\eeq
where ${\cal L}^{(0)}$ and $s^{(0)}$ are the 
Lagrangian and the BRST operator for new minimal SUGRA
and $\Phi^*_A$ denotes the antifields.
The deformed solution of the master
equation will still depend only linearly on the antifields, i.e.\ the
gauge algebra will be closed off-shell even in the deformed theory and
${\cal S}$ will take the form
\beq {\cal S}=\int d^4x\, \left({\cal L}-(s\Phi^A)\Phi^*_A\right)
\label{me6}\eeq
where ${\cal L}$ and $s\Phi^A$ will not involve antifields.

\subsection{Toy model revisited}\label{toy}

Explicit computations in SUGRA are often rather
involved. Appropriate methods that simplify the
calculations are therefore very welcome. In this section
I illustrate the technique used to derive the results of
this paper for a toy model discussed already in \cite{sugra}.
The analogous calculation in the SUGRA case is sketched in
appendix \ref{appA}.

The toy model is defined in flat four dimensional Minkowski
space by the Lagrangian
\beq
{\cal L}^{(0)}=-2\epsilon^{\mu\nu\rho\sigma}a_\mu \6_\nu t_{\rho\sigma}
-\s0 14\, F^{\mu\nu}F_{\mu\nu}\label{toy1}\eeq
where $a_\mu$ and $t_{\mu\nu}$ are the components of
an abelian gauge field and
2-form gauge potential respectively, and $F$ is
the field strength of a second abelian gauge field,
\beq F_{\mu\nu}=\6_\mu A_\nu-\6_\nu A_\mu\ .
\label{toy1a}\eeq
The toy model is evidently invariant under gauge transformations
corresponding to the following simple BRST transformations:
\beq 
\ba{lll}
s^{(0)}t_{\mu\nu}=\6_\nu Q_\mu -\6_\mu Q_\nu\ , &
s^{(0)}Q_\mu=\6_\mu Q\ ,& s^{(0)}Q=0\ ,\\
s^{(0)}a_\mu=\6_\mu c\ ,& s^{(0)}c=0\ ,& \\
s^{(0)}A_\mu=\6_\mu C\ ,& s^{(0)}C=0 &
\ea
\label{toy2}
\eeq
where $Q_\mu$, $c$ and $C$ are the ghost fields corresponding
to $t_{\mu\nu}$, $a_\mu$ and $A_\mu$ respectively, and
$Q$ is a ghost for the ghosts $Q_\mu$ (i.e.\ $Q$ has ghost number
two). Due to the closure of the gauge algebra, the
proper solution ${\cal S}^{(0)}$ of the master equation corresponding
to (\ref{toy1}) and (\ref{toy2}) is just of the form (\ref{me5}).
According to the standard rules of the field-antifield formalism,
the BRST transformations of the antifields are then
obtained from
\beq
s^{(0)}\Phi^*_A=\left({\cal S}^{(0)},\Phi^*_A\right)
=\frac {\delta^R{\cal S}^{(0)}}{\delta \Phi^A}\ .
\eeq
Consider now the following total forms (= formal sums of
local differential forms):
\bea
\4C^*&=&d^4xC^*
+\s0 16\,dx^\mu dx^\nu dx^\rho \epsilon_{\mu\nu\rho\sigma}A^{\sigma *}
+\s0 14\,dx^\mu dx^\nu \epsilon_{\mu\nu\rho\sigma}F^{\rho\sigma}\ ,
\label{toy3}\\
\4Q&=&Q+dx^\mu Q_\mu+\s0 12\,dx^\mu dx^\nu t_{\mu\nu}\ ,
\label{toy4}\\
H&=&\s0 12\, dx^\mu dx^\nu dx^\rho\6_\mu t_{\nu\rho}
\label{toy4b}\eea
where $C^*$ and $A^{\mu *}$ are the antifields of $C$ and $A_\mu$,
the differentials $dx^\mu$ are treated as Grassmann odd (anticommuting)
quantities, and 
\[ d^4x=dx^0dx^1dx^2dx^3
=-\s0 1{24}\, \epsilon_{\mu\nu\rho\sigma}
dx^\mu dx^\nu dx^\rho dx^\sigma\ .\]
It is easy to verify that $\4C^*$ and $\4Q$ satisfy
\beq 
\4s^{(0)}\4C^*=0\ ,\quad \4s^{(0)}\4Q=H
\label{toy4a}\eeq
where $\4s^{(0)}$ is the sum of $s^{(0)}$ and
the spacetime  exterior derivative $d=dx^\mu\6_\mu$,
\beq 
\4s^{(0)}=s^{(0)}+d\ .
\eeq
(\ref{toy4a}) implies evidently
\beq 
\4s^{(0)}(\4C^*\4Q)=\4C^*H=0
\label{toy4c}\eeq
where the second equality holds because $\4C^*H$ 
contains only form degrees exceeding four.
(\ref{toy4c}) decomposes of course into 
the so-called descent equations 
$s^{(0)}\omega_4+d\omega_3=0$,
$s^{(0)}\omega_3+d\omega_2=0$, 
$s^{(0)}\omega_2=0$ satisfied by the
$p$-forms  $\omega_p$ contained in $\4C^*\4Q=\sum\omega_p$
($p=2,3,4$).
In particular it thus implies that $\int \omega_4$
is $s^{(0)}$-invariant (up to the boundary term
$\int d(-\omega_3)$). Furthermore, $\int \omega_4$
is cohomological nontrivial, because
its antifield independent part 
does not vanish on-shell up to a boundary term,
and is thus a candidate first order deformation
which reads explicitly
\beq {\cal S}^{(1)}=
\int d^4x\, (-\s0 12\, F^{\mu\nu}t_{\mu\nu}
+A^{\mu *}Q_\mu+C^*Q)\ .
\label{toy5}\eeq
It is straightforward to verify that, dropping a boundary term,
\beq \left({\cal S}^{(1)},{\cal S}^{(1)}\right)
=2\int d^4x\, Q_\nu \6_\mu t^{\mu\nu}
=s^{(0)}\int d^4x\, \s0 12\, t_{\mu\nu}t^{\mu\nu}\ .
\eeq
Hence, the first three equations (\ref{me3}) are 
satisfied with ${\cal S}^{(1)}$ as in (\ref{toy5}) and
\beq 
{\cal S}^{(2)}=-\int d^4x\, \s0 12\, t_{\mu\nu}t^{\mu\nu}\ .
\eeq
Evidently ${\cal S}^{(1)}$ and ${\cal S}^{(2)}$ satisfy
\beq
\left({\cal S}^{(1)},{\cal S}^{(2)}\right)=
\left({\cal S}^{(2)},{\cal S}^{(2)}\right)=0\ .
\eeq
We thus conclude that a deformed solution of the master equation
is given by
\bea
{\cal S}&=&{\cal S}^{(0)}+g\, {\cal S}^{(1)}
              +\s0 12\, g^2{\cal S}^{(2)}
\nonumber\\
&=&\int d^4x\, \{
-2\epsilon^{\mu\nu\rho\sigma}a_\mu \6_\nu t_{\rho\sigma}
-\s0 14\, (F^{\mu\nu}+gt^{\mu\nu})(F_{\mu\nu}+gt^{\mu\nu})
\nonumber\\
& &
\phantom{\int d^4x \,\{}
+t^{\mu\nu *}(\6_\nu Q_\mu -\6_\mu Q_\nu)
- Q^{\mu *}\6_\mu Q + a^{\mu *}\6_\mu c
\nonumber\\
& &
\phantom{\int d^4x \,\{}
+ A^{\mu *}(\6_\mu C+gQ_\mu)+gC^*Q\}\ .
\label{toy6}
\eea
{}From this one reads off easily the deformed Lagrangian
and BRST transformations of the fields.

\mysection{Simplest model}
\label{new}

This section presents the result obtained by deforming
new minimal SUGRA coupled only to one
$U(1)$-gauge multiplet analogously to the toy model.
This SUGRA model has (16+16) degrees of freedom off-shell%
\footnote{We employ here the usual counting where
one counts separately the fermionic and bosonic
fields (including the auxiliary ones), subtracting respectively
the number of gauge symmetries, and adding the number of
reducibility conditions.}. The field content, including
the ghost fields, is given in the table below which also
indicates the ghost numbers ($gh$),
Grassmann parities ($\varepsilon$), dimension assignments ($dim$),
$R$-charges ($r$) and reality properties of the fields.
\[
\ba{c|c|c|c|c|l}
\Phi             &  gh(\Phi) & \varepsilon(\Phi) & dim(\Phi) & r(\Phi)\\
\hline\rule{0em}{2ex}
\viel a\mu       &       0    &     0     &  0   & 0 &
                \mbox{vierbein fields (real)}\\
C^\mu            &       1    &     1     &  -1  & 0 &
                \mbox{diffeomorphism ghosts (real)}\\
C^{ab}           &       1    &     1     &  0   & 0 &
                \mbox{Lorentz ghosts (real)}\\
\hline\rule{0em}{2ex}
\psi_\mu         &       0    &     1     &  1/2 & 1 &
                \mbox{gravitino (complex)} \\
\xi              &       1    &     0     & -1/2 & 1 &
                \mbox{SUSY ghosts (complex)}\\
\hline\rule{0em}{2ex}
t_{\mu\nu}       &       0    &     0     &    0 & 0 &
    \mbox{2-form gauge potential (real)}\\
Q_\mu            &       1    &     1     &    -1& 0 &
    \mbox{ghosts for $t_{\mu\nu}$ (real)} \\
Q                &       2    &     0     &    -2& 0 &
    \mbox{ghost for ghosts (imaginary)}\\
\hline\rule{0em}{2ex}
a_\mu            &       0    &     0     &    1 & 0 &
    \mbox{$R$-gauge field (real)}\\
c             &       1    &     1     &    0 & 0 &
    \mbox{$R$-ghost (real)}\\
\hline\rule{0em}{2ex}
A_\mu            &       0    &     0     &   1  & 0 &
                \mbox{$U(1)$ gauge field (real)}\\
\lambda          &       0    &     1     &  3/2 & 1 &
                \mbox{$U(1)$ gaugino (complex)}\\
D                &       0    &     0     &   2  & 0 &
                \mbox{scalar aux. field (real)}\\
C                &       1    &     1     &   0  & 0 &
                \mbox{$U(1)$ ghost (real)}\\
\ea
\]
The model has the following gauge symmetries:
general coordinate and local Lorentz invariance,
local N=1 SUSY, local $R$-symmetry, the
local $U(1)$ symmetry associated with $A_\mu$, and the 
reducible gauge symmetry associated with $t_{\mu\nu}$.
The corresponding BRST transformations are given explicitly below.
The action contains two separately invariant
parts. Their integrands are denoted by $e L_{grav}$ and $e L_{U(1)}$
 and occur with coefficients $M_{Pl}^2$ and $g_0^{-2}$ respectively,
where $M_{Pl}$ is the Planck mass and $g_0$ is a coupling
constant for the $U(1)$ symmetry associated with $A_\mu$
($g_0$ may be absorbed by
rescaling $A_\mu$, $\lambda$, $D$, $C$ and $g_1$). 
The deformation parameter $g_1$ has the same dimension
as $M_{Pl}^2$, i.e.\ $g_1 M_{Pl}^{-2}$ is dimensionless.
The Lagrangian reads
\bea {\cal L}&=&e\, \left(M_{Pl}^2 L_{grav}+g_0^{-2}L_{U(1)}\right)\ ,
\label{Lnew}\\
L_{grav}&=&
\s0 12\, R
-2\varepsilon^{\mu\nu\rho\sigma}
(\psi_\mu\sigma_\nu\nabla_\rho\5\psi_\sigma
-\5\psi_\mu\5\sigma_\nu\nabla_\rho\psi_\sigma)
\nonumber\\
& &
-3H_\mu H^\mu
-2\varepsilon^{\mu\nu\rho\sigma}a_\mu \6_\nu t_{\rho\sigma} \ ,
\label{Lgrav}\\
L_{U(1)}&=&
-\s0 14\, (F_{\mu\nu}+g_1\, t_{\mu\nu})(F^{\mu\nu}+g_1\, t^{\mu\nu})
+\s0 12\, D^2-\s0 18\, g_1^2
\nonumber\\
& &
-\s0 12\, i \, (\lambda\sigma^\mu\nabla_\mu\5\lambda
+\5\lambda\5\sigma^\mu\nabla_\mu\lambda)
+\s0 12\,g_1\,(i\,\lambda\sigma^\mu\5\psi_\mu-i\,\psi_\mu\sigma^\mu\5\lambda)
\nonumber\\
& &
-\s0 12\, \varepsilon^{\mu\nu\rho\sigma}(F_{\mu\nu}+g_1\, t_{\mu\nu})
(\psi_\rho\sigma_\sigma\5\lambda+\lambda\sigma_\sigma\5\psi_\rho)
+\s0 32\, \lambda\sigma^\mu\5\lambda\, H_\mu
\nonumber\\
& &
+\psi_\mu\sigma^{\mu\nu}\psi_\nu\5\lambda\5\lambda
+\5\psi_\mu\5\sigma^{\mu\nu}\5\psi_\nu \lambda \lambda 
\label{LU1}
\eea
with
\bea
e&=&\det (\viel a\mu)\ ,
\\
\varepsilon^{\mu\nu\rho\sigma}&=&\Viel a\mu\cdots\Viel d\sigma
\varepsilon^{abcd}=e^{-1}\epsilon^{\mu\nu\rho\sigma}\quad
(\epsilon^{0123}=1)\ ,
\\
R&=&2\Viel a\nu \Viel b\mu 
(\6_{[\mu}\spin {ab}{\nu]}-\spin {ca}{[\mu}\spin b{\nu]c})\ , 
\label{Riemann}\\
H^\mu&=&\varepsilon^{\mu\nu\rho\sigma}(\s0 12\, \6_{\nu}t_{\rho\sigma}
+i\, \psi_{\nu}\sigma_\rho\5\psi_{\sigma})\ ,
\label{H}\\
F_{\mu\nu}&=&2(\6_{[\mu} A_{\nu]}
     +i\, \lambda\sigma_{[\mu}\5\psi_{\nu]}
               +i\, \psi_{[\mu}\sigma_{\nu]}\5\lambda) \ ,
\label{F}\\
\nabla_\mu\psi_\nu&=&\6_\mu\psi_\nu
-\s0 12\,\spin {ab}\mu\psi_\nu\sigma_{ab}-i\, a_\mu\psi_\nu\ ,
\label{nablapsi}\\
\nabla_\mu\lambda&=&
\6_\mu\lambda
-\s0 12\,\spin {ab}\mu\lambda\sigma_{ab}-i\, a_\mu\lambda \ .
\label{nablala}
\eea
where $\Viel a\mu$ and $\spin {ab}\mu$  
denote the components of the inverse
vielbein and of the standard gravitino dependent 
spin connection respectively, 
\bea 
& &\Viel a\mu\viel a\nu=\delta_\nu^\mu\ ,
\quad \Viel a\mu\viel b\mu=\delta_a^b\ ,
\\
& &\spin {ab}\mu = E^{a\nu}E^{b\rho}(\omega_{[\mu\nu]\rho}
-\omega_{[\nu\rho]\mu}+\omega_{[\rho\mu]\nu})\ ,
\nonumber\\
& &\omega_{[\mu\nu]\rho}=e_{\rho a}\6_{[\mu}\viel a{\nu]}
-i\psi_{\mu}\sigma_\rho\5\psi_{\nu}
+i\psi_{\nu}\sigma_\rho\5\psi_{\mu}\ .
\label{spin}
\eea

The deformed BRST transformations of the classical fields are:
\bea
 s\viel a\mu &=& C^\nu\6_\nu\viel a\mu+(\6_\mu C^\nu)\viel a\nu
+C_b{}^a\viel b\mu
\nonumber\\
& &
+2i(\xi\sigma^a\5\psi_\mu-\psi_\mu\sigma^a\5\xi)\ ,
\label{sviel}\\
s\psi_\mu&=&C^\nu\6_\nu\psi_\mu+(\6_\mu C^\nu)\psi_\nu
+\s0 12\, C^{ab}\psi_\mu\sigma_{ab}+i\, c\,  \psi_\mu
\nonumber\\
& &+\6_\mu\xi-\s0 12\,\spin {ab}\mu\xi\sigma_{ab}-i\, a_\mu\xi
-i\, \xi H_\mu-i\, \xi\sigma_{\mu\nu} H^\nu\ ,
\label{spsi}\\
st_{\mu\nu}&=&C^\rho\6_\rho t_{\mu\nu}
+(\6_\mu C^\rho)\, t_{\rho\nu}
+(\6_\nu C^\rho)\, t_{\mu\rho}
+\6_\nu Q_\mu-\6_\mu Q_\nu\nonumber\\
& &-i\,(\xi\sigma_\mu\5\psi_\nu-\xi\sigma_\nu\5\psi_\mu
+\psi_\mu\sigma_\nu\5\xi-\psi_\nu\sigma_\mu\5\xi)\ ,
\label{st}\\
sa_\mu&=&C^\nu\6_\nu a_\mu+(\6_\mu C^\nu)a_\nu
+\6_\mu c  
\nonumber\\
& &+\xi\sigma_\mu \5S+S\sigma_\mu \5\xi\ ,
\label{sa}\\
sA_\mu&=&\6_\mu C+C^\nu\6_\nu A_\mu+(\6_\mu C^\nu)A_\nu
\nonumber\\
& &
-i\, \xi\sigma_\mu\5\lambda
+i\, \lambda\sigma_\mu\5\xi+g_1\, Q_\mu\ ,
\label{sA}\\
s\lambda&=&C^\mu\6_\mu\lambda+\s0 12\, C^{ab}\lambda\sigma_{ab}
+i\, c\,  \lambda
\nonumber\\
& &
+\xi\, (\s0 12\, g_1-iD)
-\xi\sigma^{\mu\nu}(F_{\mu\nu}+g_1\, t_{\mu\nu})\ ,
\label{slambda}\\
sD&=&C^\mu\6_\mu D
\nonumber\\
& & 
+\xi \sigma^\mu \left[\nabla_\mu\5\lambda-\5\psi_\mu (iD+\s0 12\, g_1)
-\5\sigma^{\nu\rho}\5\psi_\mu (F_{\nu\rho}+g_1\, t_{\nu\rho})
\right]
\nonumber\\
& &+\left[(\nabla_\mu\lambda)+(iD-\s0 12\, g_1)\psi_\mu 
+ (F_{\nu\rho}+g_1\, t_{\nu\rho})\psi_\mu\sigma^{\nu\rho}
\right]\sigma^\mu\5\xi
\nonumber\\
& &
+\s0 32\, i\, (\xi \sigma^\mu\5\lambda -\lambda\sigma^\mu\5\xi)H_\mu
\label{sD}\eea
where $S^\alpha$ and $\5S^\da$ which occur in (\ref{sa})
are the spin-$\frac 12$ parts of the super-covariant
gravitino field strengths,
\beq
S=2(\nabla_\mu\psi_\nu)\sigma^{\mu\nu}
+\s0 32\, i\, \psi_\mu H^\mu\ ,\quad
\5S=-2\5\sigma^{\mu\nu}\nabla_\mu\5\psi_\nu
-\s0 32\, i\, \5\psi_\mu H^\mu\ .
\label{S}\eeq
The BRST transformation of the ghosts and the ghost for ghosts
are
\bea
sC^\mu&=&C^\nu\6_\nu C^\mu+2i\,\xi\sigma^\mu\5\xi\ ,
\label{sCdiff}\\
s\xi&=&C^\mu\6_\mu \xi
+\s0 12C^{ab}\xi\sigma_{ab}
+i\, c\,  \xi
-2i\, \xi\sigma^\mu\5\xi\,\psi_\mu\ ,\label{sxi}\\
sC^{ab}&=&C^\mu\6_\mu C^{ab}
+C^{ca}{C_c}^b-2i\, \xi\sigma^\mu\5\xi\,\spin {ab}\mu
+2i\,\varepsilon^{abcd}\xi\sigma_c\5\xi \, H_d\ ,
\label{sCLor}\\
sc  &=&C^\mu\6_\mu c  
-2i\, \xi\sigma^\mu\5\xi\, a_\mu\ ,
\label{sc}\\
sC&=&C^\mu\6_\mu C 
-2i\, \xi\sigma^\mu\5\xi\, A_\mu-g_1\, Q\ ,
\label{sC}\\
sQ_\mu&=&\6_\mu Q+C^\nu\6_\nu Q_\mu+(\6_\mu C^\nu)Q_\nu
-2i\, \xi\sigma^\nu\5\xi\, t_{\mu\nu}-i\, \xi\sigma_\mu\5\xi\ ,
\label{sQmu}\\
sQ&=&C^\mu\6_\mu Q-2i\, \xi\sigma^\mu\5\xi \, Q_\mu\ .
\label{sQ}
\eea
In the formulas 
(\ref{sviel})--(\ref{sQ}) 
the spinor indices of 
$\psi_\mu$, $\5\psi_\mu$, $\lambda$, $\5\lambda$, $S$, $\5S$,
$\xi$ and $\5\xi$ are everywhere upstairs. The above 
BRST transformations are off-shell nilpotent, 
\beq s^2\Phi=0\quad\forall\Phi\ .\eeq
As $s\Phi$ does
not involve antifields, the gauge algebra 
closes off-shell, as promised.

Notice that $g_1$ appears only in $e L_{U(1)}$ and in the 
BRST transformations of $A_\mu$, $\lambda$, $D$ and $C$.
Notice also that the deformation introduces
a  cosmological constant and spontaneous
SUSY breaking which are not present for $g_1=0$. 
$\lambda$ is the Goldstone
fermion for the broken SUSY. The cosmological constant can
be removed when matter multiplets are included, 
cf.\ section \ref{matter}.
Notice that (\ref{slambda})
looks as if the $D$-field has obtained a constant
{\em imaginary} part through the deformation.

For later purpose I note that the equations of motion for
$\psi_\mu$, $a_\mu$ and $t_{\mu\nu}$ imply the following
on-shell equalities (indicated by $\approx$):
\bea
S^\alpha &\approx& - g_1(2g_0M_{Pl})^{-2}\lambda^\alpha\ ,
\label{eomS}\\
H^\mu &\approx& (2g_0M_{Pl})^{-2}\lambda\sigma^\mu\5\lambda\ , 
\label{eomH}\\
2\6_{[\mu} a_{\nu]} &\approx&
g_1(2g_0M_{Pl})^{-2}
\{2\lambda\sigma_{[\mu}\5\psi_{\nu]}
-2\psi_{[\mu}\sigma_{\nu]}\5\lambda
+\s0 12\, \varepsilon_{\mu\nu\rho\sigma}
(F^{\rho\sigma}+g_1t^{\rho\sigma})\} .
\label{eoma}
\eea

\mysection{Relation to old minimal SUGRA and duality}\label{disc}

It will now be shown that the deformed model of section \ref{new} 
is classically (on-shell)
equivalent to old minimal SUGRA with
local $R$-symmetry for all nonvanishing values
of $g_1$. In contrast, 
$g_1=0$ gives of course new minimal
SUGRA coupled to a $U(1)$ gauge multiplet.

The reason is that for $g_1\neq 0$ both the deformed action and 
gauge resp.\ BRST transformations depend on the $t_{\mu\nu}$ and
$A_\mu$ only via the combinations $g_1 t_{\mu\nu}+
\6_\mu A_\nu-\6_\nu A_\mu$. 
For $g_1\neq 0$, we can therefore introduce
these combinations as new elementary fields instead of the $t_{\mu\nu}$
(in fact we will use a slightly different choice below).
These new fields become auxiliary and can be eliminated
algebraically after
redefining also the $R$-gauge field appropriately. After
elimination of the auxiliary fields,
it becomes evident that the deformed model is (for $g_1\neq 0$)
indeed classically equivalent
to old minimal SUGRA with local $R$-symmetry. Note however that
the auxiliary field content differs from that of old minimal SUGRA:
instead of a real vector field and a complex scalar field,
the deformed model contains after the field redefinitions
an auxiliary real antisymmetric tensor field. It should be
kept in mind that the field redefinitions make sense only for
$g_1\neq 0$, i.e.\ the redefined fields cannot be used to
describe the complete model, in contrast to the original
fields used in section \ref{new}.

Convenient field redefinitions are
\bea
b_{\mu\nu} &=& g_0^{-1}g_R(g_1 t_{\mu\nu}+F_{\mu\nu})\ ,
\label{b}\\
\7a_\mu &=& a_\mu+\s0 34\, H_\mu
-\s0 3{16}\, (g_0M_{Pl})^{-2}\lambda\sigma_\mu\5\lambda\ ,
\label{7a}\\
\7\lambda &=& i g_0^{-1}g_R\lambda\ ,
\label{hatla}\\
\7D &=& g_0^{-1}g_R D
\label{hatD}
\eea
with $F_{\mu\nu}$ and $H_\mu$ as in (\ref{F}) and (\ref{H})
respectively, and
\beq g_R=\frac{g_1}{4g_0M_{Pl}^2}\ .
\label{gR}\eeq
$g_R$ is the dimensionless $R$-coupling constant in the
deformed theory. This becomes clear when one writes the
deformed Lagrangian (\ref{Lnew}) in terms of the redefined fields:
\bea
{\cal L}_{(g_1\neq 0)}&=& e\,( M_{Pl}^2L_1+g_R^{-2} L_2+g_R^{-2} L_3)
\label{Lredef}\\
L_1&=&
\s0 12 R-2\,\varepsilon^{\mu\nu\rho\sigma}
(\psi_\mu\sigma_\nu\7\nabla_\rho\5\psi_\sigma
-\5\psi_\mu\5\sigma_\nu\7\nabla_\rho\psi_\sigma)
\nonumber\\
& &
+2 (\7\lambda\sigma^\mu\5\psi_\mu
      +\psi_\mu\sigma^\mu\5{\7\lambda})-2(g_R M_{Pl})^2\ ,
\label{L1}\\
L_2&=&
-\s0 i2\7\lambda\sigma^\mu\7\nabla_\mu\5{\7\lambda}
+\s0 12\varepsilon^{\mu\nu\rho\sigma}
       \7F_{\mu\nu}\7\lambda\sigma_\rho\5\psi_\sigma
\nonumber\\
& &
+\5\psi_\mu\5\sigma^{\mu\nu}\5\psi_\nu \7\lambda \7\lambda
+\s0 3{16}\, (g_R M_{Pl})^{-2}
\7\lambda\7\lambda\,\5{\7\lambda}\5{\7\lambda}+c.c.\ ,
\label{L2}\\
L_3&=&\s0 12 \7D^2-\s0 14\, b_{\mu\nu}(b^{\mu\nu}
+\varepsilon^{\mu\nu\rho\sigma}\7F_{\rho\sigma})
\label{L3}\eea
with
\bea
\7F_{\mu\nu} &=& 2(\6_{[\mu} \7a_{\nu]}
     +i\, \7\lambda\sigma_{[\mu}\5\psi_{\nu]}
               +i\, \psi_{[\mu}\sigma_{\nu]}\5{\7\lambda}) \ ,
\label{hatF2}\\
\7\nabla_\mu \psi_\nu &=& 
\6_\mu\psi_\nu-\s0 12\,\spin {ab}\mu \psi_\nu\sigma_{ab}
                 -i \7a_\mu \psi_\nu\ ,
\label{7nabla}\\
\7\nabla_\mu \7\lambda &=& 
\6_\mu\7\lambda-\s0 12\,\spin {ab}\mu \7\lambda\sigma_{ab}
                 -i \7a_\mu \7\lambda\ .
\eea
As promised, $b_{\mu\nu}$ is an auxiliary field. Its
classical equation of motion reads
\beq b^{\mu\nu}\approx
-\s0 12\, \varepsilon^{\mu\nu\rho\sigma}\7F_{\rho\sigma}\ .
\label{eom}\eeq
Upon elimination of $b_{\mu\nu}$ and $\7D$, $L_3$ becomes
$(-\s0 14\7F_{\mu\nu}\7F^{\mu\nu})$ and
the complete action turns indeed into the one 
of old minimal SUGRA with local $R$-symmetry  
(see e.g.\ \cite{sugra})
after eliminating the auxiliary fields there too.

The BRST transformations for the redefined fields
can be obtained from the formulas of section \ref{new}.
For instance one gets
\bea
s\7\lambda&=&C^\mu\6_\mu\7\lambda
+\s0 12\, C^{ab}\7\lambda\sigma_{ab}
+ic\,  \7\lambda
\nonumber\\
& &
+\xi\, (\7D+2ig_R^2M_{Pl}^2)-i\xi\sigma^{\mu\nu}b_{\mu\nu}\ .
\label{shatla}
\eea
$\7\lambda$ plays for $g_1\neq 0$ the
role of the gaugino for $R$-transformations.
This is read off from the above formulas, taking
into account (\ref{eom}) and (\ref{eomS}). Note that
the latter reads
in terms of the redefined fields just $S\approx i\7\lambda$
which ``explains'' why $\7\lambda$ turns into the $R$-gaugino.
Namely $S$ plays in fact in new minimal SUGRA the role
of a composite $R$-gaugino (cf.\ e.g.\ appendix
B of \cite{sugra}).

Let me now briefly discuss the electromagnetic
duality mentioned in the introduction. It
relates, for $g_1\neq 0$, the $U(1)$-symmetry 
associated with $A_\mu$ to the $R$-symmetry and
is established by (\ref{eom}) (resp.\ by (\ref{eoma})).
The latter shows indeed that $b_{\mu\nu}$, which contains
the super-covariant $U(1)$ field strength $F_{\mu\nu}$,
is (on-shell) dual to the super-covariant $R$-field strength
$\7F_{\mu\nu}$. Moreover, we have just seen that $\lambda$,
which is for $g_1=0$ the gaugino associated with 
$A_\mu$, turns for $g_1\neq 0$ on-shell
into the $R$-gaugino. Last but not least, it
is striking that the
$R$-coupling constant $g_R$ is proportional to the
inverse $U(1)$-coupling constant $g_0$, cf.\ (\ref{gR}).

In this context it is worthwhile to recall that
for $g_1=0$, i.e.\ in new minimal SUGRA, 
there is no $R$-coupling constant in the usual 
sense, which reflects that one cannot switch off the $R$-symmetry
in new minimal SUGRA without
switching off simultaneously SUSY. 
Furthermore, on-shell the $R$-gauge field $a_\mu$ is for $g_1=0$ 
pure gauge, at least locally, cf.\ (\ref{eoma}).
Hence, for $g_1=0$ this gauge field does not carry 
local physical degrees of freedom, in contrast to $A_\mu$.
Turning on $g_1$, $A_\mu$ disappears effectively
through the above field redefinitions, and transfers
its degrees of freedom via the duality to the 
(redefined) $R$-gauge field.

\mysection{Extensions}\label{gen}

\subsection{Inclusion of a Fayet-Iliopoulos term}

We will now discuss extensions of the simple model
discussed in section \ref{new}. 
First we include a Fayet-Iliopoulos term
for the $U(1)$ symmetry associated with $A_\mu$. This
term is introduced by a second deformation of the 
model and we denote the corresponding deformation parameter
by $g_2$ (it has the same dimension as $g_1$ and $M_{Pl}^2$). 
It turns out that this deformation does not
cause further modifications of the gauge resp.\ BRST transformations
and adds only the following term to the Lagrangian which is
separately invariant under the BRST transformations given 
in section \ref{new} up to a total derivative:
\bea
{\cal L}_{FI}&=&
g_2\, e\, 
(D+\lambda\sigma^\mu\5\psi_\mu+\psi_\mu\sigma^\mu\5\lambda
\nonumber\\
& &
+\varepsilon^{\mu\nu\rho\sigma}A_\mu \6_\nu t_{\rho\sigma}
+\s0 14\, g_1\,\varepsilon^{\mu\nu\rho\sigma} t_{\mu\nu} t_{\rho\sigma})\ .
\label{LFI}
\eea
Evidently this term provides for all
values of $g_1$ an additional contribution to the
cosmological constant.
Repeating the discussion of section \ref{disc}, one
finds that the inclusion of (\ref{LFI})
results for $g_1\neq 0$ again in a model which is classically
equivalent to old minimal SUGRA with local $R$-symmetry.
The only differences are shifts of the cosmological constant,
the $R$-coupling constant
and the SUSY breaking parameter, and the occurrence
of a  theta term for the $R$-symmetry. The latter is
of course classically irrelevant. 
The $R$-coupling constant is now
\beq g'_R=\frac{\sqrt{(g_1/g_0)^2+(2g_2g_0)^2}}{4M_{pl}^2}\ .
\label{newgR}\eeq
It is the modulus of a complex parameter
\[ z=\frac{2g_2g_0+ig_1/g_0}{4M_{pl}^2}=g'_R e^{i\theta} \]
whose phase appears in the coefficient of the
above-mentioned theta term for the $R$-symmetry according to 
\[ -\s0 12\, (g'_R)^{-2} \cot \theta\, \varepsilon^{\mu\nu\rho\sigma}
\6_\mu \7a_\nu \6_\rho \7a_\sigma \]
with $\7a_\mu$ as in (\ref{7a}) (note that
$\theta=0$ is excluded here as it corresponds to $g_1=0$). 
(\ref{7a}) and 
field redefinitions analogous to (\ref{b}), (\ref{hatla})
and (\ref{hatD}), namely 
\bea
b'_{\mu\nu} &=& g'_Rg_0^{-1}(g_1 t_{\mu\nu}+F_{\mu\nu})\ ,
\label{b'}\\
\7\lambda' &=& g'_R g_0^{-1} e^{i\theta}\lambda\ ,
\label{hatla'}\\
\7D' &=& g'_R (g_0^{-1}D+g_0g_2)\ ,
\label{hatD'}
\eea
establish for $g_1\neq 0$ again the classical 
equivalence of the model based on the
sum of (\ref{Lnew}) and (\ref{LFI}) to old minimal
SUGRA with local $R$-symmetry.
In particular $b'_{\mu\nu} $ becomes again an auxiliary field,
and after eliminating it one arrives in fact at an action of exactly
the same form as in section \ref{disc} after eliminating
$b_{\mu\nu} $ there. The only differences are
that $\7\lambda$, $\7D$ and $g_R$ are replaced by
$\7\lambda'$, $\7D'$ and $g'_R$ respectively, and that the
above theta term appears.

Notice that the appearance of the $R$-coupling constant
and the coefficient of the theta term in a single complex
parameter $z$ is reminiscent of similar (though
somewhat different) relations in other globally
and locally supersymmetric gauge theories.

\subsection{Coupling to matter and further gauge multiplets}
\label{matter}

The inclusion of matter or further gauge multiplets is
completely straightforward. Indeed, recall that the deformation
modifies only the BRST transformations of $A_\mu$, $\lambda$,
$D$ and $C$, but not the gauge transformation of any other field.
Hence, any term involving matter or further gauge multiplets
which is gauge invariant in new minimal SUGRA will be
gauge invariant in the deformed theory too, provided it
does not involve $A_\mu$, $\lambda$ or $D$. Of course
this requires in particular that all the matter fields
transform trivially under the $U(1)$ symmetry associated
with $A_\mu$. The latter requirement 
however is just a necessary prerequisite for the
existence of the deformation in presence of matter fields,
as pointed out in \cite{sugra}, and must thus be imposed
anyhow.

In particular,
gauge invariant contributions to the action containing 
the kinetic terms for Yang--Mills multiplets, as well as 
the kinetic and superpotential terms
for chiral matter multiplets are thus
exactly the same as in new minimal SUGRA.
Therefore the cosmological constant implemented
by the deformation can be removed by the usual mechanism when
matter multiplets are included,
cf.\ \cite{cos,dual2} and, more recently, \cite{cris}. 
Furthermore, the standard kinetic and superpotential terms
for matter and gauge multiplets depend 
on $t_{\mu\nu}$ only through
its super-covariant field strength $H_\mu$ given in
(\ref{H}). As a shift of $t_{\mu\nu}$ by $\6_\mu A_\nu
-\6_\nu A_\mu$ drops out in $H_\mu$, one concludes
again that the complete action depends for $g_1\neq 0$
on $t_{\mu\nu}$ and $A_\mu$ only through $b_{\mu\nu}$.

\mysection{Anomalies}\label{anos}

It is well-known that the presence of
a classical $R$-gauge symmetry can
lead to chiral anomalies%
\footnote{Besides the ``pure'' $R$-anomaly, these
are also ``mixed'' anomalies, 
such as mixed gravitational and $R$-anomalies.}, 
at least in old minimal SUGRA.
As the latter is classically equivalent to
our model for $g_1\neq 0$, one expects
that the $R$-symmetry will implement chiral anomalies
in our model too, unless all the contributions to
these anomalies (by the gravitino, gauginos and matter
fermions) cancel. Furthermore, the same argument suggests that
these anomaly cancellation conditions coincide with those in
old minimal SUGRA analysed recently in \cite{cris}.

As anomalies correspond
to BRST cohomology classes at ghost number one, it is
therefore instructive to look for representatives of 
such cohomology classes which can correspond to
chiral anomalies associated with the $R$-symmetry in our model.
In particular ``the'' (pure) $R$-anomaly is of interest
in this context, as it has some unusual features as compared
to chiral anomalies in more standard theories. 
Namely, inspired by the familiar representatives of
chiral anomalies, 
one might expect that this anomaly is represented by a 
BRST-invariant functional of the
form $\int c \, da\, da+\mbox{``more''}$ where 
$a=dx^\mu a_\mu$  is the
$R$-connection and $\mbox{``more''}$ indicates
terms to be chosen such that the complete expression
is invariant under the BRST transformations 
given in section \ref{new}.
However, in our case $\int c \, da\, da$ vanishes
on-shell up to terms of higher order in $g_1$, cf.\ (\ref{eoma}). 
Therefore this term can
be removed from any BRST-invariant functional by
subtracting a cohomologically trivial 
(BRST-exact) local functional.
(\ref{eoma}) now suggests an alternative expression 
representing the $R$-anomaly in our model, namely
$\int c\,  B^2+\mbox{``more''}$ where
$B$ is the 2-form corresponding to (\ref{b}). As shown
in appendix \ref{appB}, one finds indeed
a corresponding BRST-invariant functional which 
reads in complete form
\bea 
{\cal A}&=&
    \int d^4x\, e  \left\{
    \s0 14\, c\, \varepsilon^{\mu\nu\rho\sigma}(F_{\mu\nu}+g_1t_{\mu\nu})
    (F_{\rho\sigma}+g_1t_{\rho\sigma})
\right.
\nonumber\\
& &
\phantom{\int d^4x\, e  \left\{\right.}
   +i\, \varepsilon^{\mu\nu\rho\sigma}
   (\lambda\sigma_\mu\5\xi-\xi\sigma_\mu\5\lambda )\, a_\nu\,
   (F_{\rho\sigma}+g_1t_{\rho\sigma})
\nonumber\\
& &
\phantom{\int d^4x\, e  \left\{\right.}
  +\xi S\, \5\lambda \5\lambda-2\xi\lambda\, \5S \5\lambda
  +\5S\5\xi\, \lambda \lambda-2\5\lambda\5\xi\, S \lambda
\nonumber\\
& &
\phantom{\int d^4x\, e  \left\{\right.}
  -g_1c\, (D+\lambda\sigma^\mu\5\psi_\mu+\psi_\mu\sigma^\mu\5\lambda)
\nonumber\\
& &
\phantom{\int d^4x\, e  \left\{\right.}
\left.
  +g_1(\xi\sigma^\mu\5\lambda +\lambda\sigma^\mu\5\xi)\, a_\mu
\right\}
\label{an1}\eea
with $F_{\mu\nu}$, $S$ and $\5S$ as in (\ref{F}) and
(\ref{S}). Note that (\ref{an1}) does not
depend on antifields and would thus provide
a universal representative of the $R$-anomaly which
does not change even in presence of
matter or further gauge multiplets.
Nevertheless it is instructive to
cast it in a different but equivalent form. 
Namely, adding to it $sX$ with
\[ X=\int d^4x\, e\, \varepsilon^{\mu\nu\rho\sigma}
a_\mu A_\nu (\6_\rho A_\sigma+g_1t_{\rho\sigma})\]
and dropping total derivatives in the
integrand, one gets (cf.\ appendix \ref{appB})
\bea
\lefteqn{
{\cal A}+sX\approx g_1\int d^4x\, e \{
\varepsilon^{\mu\nu\rho\sigma}(c\,  A_\mu-C\, a_\mu)\, \6_\nu t_{\rho\sigma}
}
\nonumber\\
& &
-\s0 18\,(g_0M_{Pl})^{-2}C F_{\mu\nu}F^{\mu\nu}
+\mbox{terms with fermions}\}+O(g_1^2)
\label{an2}\eea
where we have restricted ourselves to the simple model described in
section \ref{new}. (\ref{an2}) shows that ${\cal A}+sX$ contains on-shell
at lowest order in $g_1$ a 
linear combination of the two terms
$e\, \varepsilon^{\mu\nu\rho\sigma}(c\,  A_\mu-C a_\mu)\6_\nu t_{\rho\sigma}$
and $e\, C F_{\mu\nu}F^{\mu\nu}$ which indeed give rise to
nontrivial and inequivalent BRST-invariant functionals
in new minimal SUGRA, cf.\ \cite{sugra}, section 9. 
This signals (though it does not prove)
that ${\cal A}$ is cohomologically nontrivial in the deformed theory
and can thus indeed represent an anomaly.

\section*{Acknowledgements}
I appreciated informative conversations with Cristina
Manuel, especially about $R$-anomalies.
This work was supported by a grant of the
Spanish ministry of education and science (MEC), and
by the Commission of the European Communities contract
CHRX-CT93-0362(04).

\appendix

\mysection{Details of the calculations}

\subsection{Deformation}\label{appA}

In the following $s^{(0)}$ denotes the BRST operator
in new minimal SUGRA obtained from (\ref{sviel})--(\ref{sQ}) 
for $g_1=0$.
First one computes the analogue of (\ref{toy3}) in 
new minimal SUGRA. This means simply to compute
a solution of the descent
equations corresponding to $d^4x C^*$. The existence
of such a solution follows from $s^{(0)}C^*=\6_\mu
(C^\mu C^*-A^{\mu *})$. The result can be conveniently
expressed in terms of the following quantities which
are useful also for the computation itself:
\bea
\7D^* &=& e^{-1}D^*\ ,
\label{app1}\\
\7\lambda^*_\alpha &=& e^{-1}(\lambda^*
   +\sigma^\mu\5\psi_\mu D^*)_\alpha\ ,
\label{app2}\\
\7{\5\lambda}{}^*_\da &=& e^{-1}(\5\lambda^*
   -\psi_\mu\sigma^\mu D^*)_\da\ ,
\label{app3}\\
\7A^{a*} &=&\viel a\mu ( e^{-1}A^{\mu *}
       +2\7\lambda^*\sigma^{\mu\nu}\psi_\nu
       -2\5\psi_\nu\5\sigma^{\mu\nu}\7{\5\lambda}{}^*
       +2i\varepsilon^{\mu\nu\rho\sigma}\psi_\nu\sigma_\rho\5\psi_\sigma \7D^*)\ ,
\label{app4}\\
\7C^* &=& e^{-1} C^*\ ,
\label{app5}\\
\4\xi^a&=&(C^\mu+dx^\mu)\viel a\mu\ ,
\label{app6}\\
\4\xi^\alpha&=&\xi^\alpha+(C^\mu+dx^\mu)\RS \alpha\mu\ ,
\label{app7}\\
\4\xi^\da&=&\5\xi^\da-(C^\mu+dx^\mu)\bRS \da\mu\ .
\label{app8}\eea
Without going into details I note that (\ref{app1})--(\ref{app5})
are super-covariant (combinations of) antifields 
in the sense that their
BRST transformations do not contain derivatives of ghosts.
They correspond to the super-covariant version of the
equations of motion.
(\ref{app6})--(\ref{app8}) are ``generalized connections''
used already in \cite{sugra1,sugra}. In terms of these
quantitites, the sought analogue of (\ref{toy3}) reads
\bea
\4C^*
&=& \Xi \, \7C^*+\s0 16\,\4\xi^a\4\xi^b\4\xi^c\varepsilon_{abcd}\7A^{d*}
+\s0 12\, i\, \4\xi^a ( \7\lambda^*\sigma_a\5\vartheta
+\vartheta\sigma_a\7{\5\lambda}{}^*)
\nonumber\\
& &
+2i\, \Theta \7D^*
+\s0 14\, \4\xi^a\4\xi^b\varepsilon_{abcd} F^{cd}
+\vartheta\lambda-\5\vartheta\5\lambda
\label{app9}\eea
where $F_{ab}=\Viel a\mu\Viel b\nu F_{\mu\nu}$ with
$F_{\mu\nu}$ as in (\ref{F}) and
\bea
& &\Xi=-\s0 1{24}\,\4\xi^a\4\xi^b\4\xi^c\4\xi^d\varepsilon_{abcd}\ ,
\label{Xi}\\
& &\vartheta^\alpha = \4\xi_\da\4\xi^{\da\alpha}\ ,\quad
\5\vartheta^\da = \4\xi^{\da\alpha}\4\xi_\alpha\ ,\quad
\Theta =\4\xi_\alpha\4\xi^{\da\alpha}\4\xi_\da
\label{ths}
\eea
with $\4\xi^{\da\alpha}=\4\xi^a\5\sigma_a{}^{\da\alpha}$.
The SUGRA-analogues of (\ref{toy4}) and (\ref{toy4b}) read
\bea
\4Q&=&Q+(C^\mu+dx^\mu) Q_\mu+
\s0 12\,(C^\mu+dx^\mu)(C^\nu+dx^\nu)\, t_{\mu\nu}\ ,
\label{app10}\\
{\cal H}&=&\s0 16\, \4\xi^a\4\xi^b\4\xi^c\varepsilon_{abcd}H^d+i\,\Theta
\label{app11}\eea
with $H^a=H^\mu\viel a\mu$, $H^\mu$ as in (\ref{H}).
$\4C^*$, $\4Q$ and ${\cal H}$ satisfy identities analogous to
(\ref{toy4a}),
\beq
\4s^{(0)}\4C^*=0\ , \quad \4s^{(0)}\4Q={\cal H}
\label{app12}
\eeq
with $\4s^{(0)}=s^{(0)}+d$.
However, in contrast to (\ref{toy4c}), 
$\4s^{(0)}(\4C^*\4Q)=\4C^*{\cal H}$ does {\em not} vanish
in the SUGRA case because it contains pieces with degrees
2, 3 and 4 in the $\4\xi^a$ (the $\4\xi^a$ play a part analogous
to the differentials in the toy model).
Therefore we have to seek an $\4s^{(0)}$-invariant
completion of $\4C^*\4Q$. The existence of this completion
was proved in \cite{sugra}. It can be
efficiently computed using a technique described in
section 4 of \cite{sugra1}. The result is the following
$\4s^{(0)}$-invariant total form:
\beq \omega=\4Q\4C^*
     +\s0 12\, i\, (\eta\lambda-\5\eta\5\lambda)
     +\s0 12\, \Xi\, (\4\xi^\alpha\7\lambda^*_\alpha
     -\4\xi^\da\7\lambda^*_\da)
\label{omega}\eeq
where
\beq \eta^\alpha=-\s0 i6\, \vartheta^\beta
\4\xi_{\beta\dbe}\4\xi^{\dbe\alpha}\ ,\quad
\5\eta^\da=\s0 i6\, \4\xi^{\da\beta}\4\xi_{\beta\dbe}
\5\vartheta^\dbe\ .
\label{eta}\eeq
The volume form $\omega_4$ contained in $\omega$ provides
${\cal S}^{(1)}=\int\omega_4$. Explicitly one gets
\bea 
{\cal S}^{(1)}&=&\int d^4x\, e\, K\ ,
\\
K &=& -\s0 12 \, t_{\mu\nu}F^{\mu\nu}
+ \s0 12\, t_{\mu\nu}\varepsilon^{\mu\nu\rho\sigma}
(\lambda\sigma_\rho\5\psi_\sigma+\psi_\sigma\sigma_\rho\5\lambda)
+\s0 i2\,(\lambda\sigma^\mu\5\psi_\mu-\psi_\mu\sigma^\mu\5\lambda)
\nonumber\\
& &+QC^*-Q_\mu A^{\mu*}
+(\s0 12\, \xi- t_{\mu\nu}\xi\sigma^{\mu\nu})\lambda^*
-\5\lambda^*(\s0 12\, \5\xi+ \5\sigma^{\mu\nu}\5\xi\, t_{\mu\nu} )
\nonumber\\
& &
+(\s0 12\, \xi\sigma^\mu\5\psi_\mu+\s0 12\, \psi_\mu\sigma^\mu\5\xi
+ t_{\mu\nu}\xi\sigma^\rho\5\sigma^{\mu\nu}\5\psi_\rho 
- t_{\mu\nu}\psi_\rho\sigma^{\mu\nu}\sigma^\rho\5\xi)D^*\ .
\label{S1}\eea
As in the case of the toy model, it is now easy to complete
the computation of the deformation. The result is 
\beq {\cal S}={\cal S}^{(0)}+g_1 {\cal S}^{(1)}
              +\s0 12\, g_1^2{\cal S}^{(2)}
\eeq
with
\beq  
{\cal S}^{(2)}=-\int d^4x\, e\, (\s0 12\, t_{\mu\nu} t^{\mu\nu}+\s0 14)\ .
\label{S2}\eeq
{\em Technical remark:}

I used the convention that the set $\{\Phi^A\}$ contains
$\lambda$ and $\5\lambda$ with spinor indices {\em downstairs}.
Hence, the sum $(-s\Phi^A)\Phi^*_A$ in (\ref{me6}) contains
for instance $(-s\lambda_\alpha)\lambda^{\alpha *}=
+(s\lambda^\alpha)\lambda^*_\alpha$, and (\ref{slambda}) and
(\ref{S1}) are consistent because the spinor indices of
$s\lambda$ are {\em upstairs} in both formulas.

Furthermore I remark that ${\cal S}$ is required to be real.
In the conventions of \cite{sugra} used here, this corresponds
for instance to the following reality properties of the 
antifields of $\lambda$, $\5\lambda$, $D$ and $A_\mu$:
\[
\5\lambda^*=-\overline{\lambda^*}\ , \quad
D^*=-\overline{D^*}\ , \quad
A^{\mu *}=-\overline{A^{\mu *}}\ .
\]

\subsection{Candidate anomaly}\label{appB}

The results given in section \ref{anos} can be derived analogously,
using now $\4s=s+d$ rather than $\4s^{(0)}$ and
in addition the following total forms:
\bea
\4C &=& C+(C^\mu+dx^\mu)\, A_\mu\ ,
\\
\4c  &=& c +(C^\mu+dx^\mu)\, a_\mu\ ,
\\
{\cal B} &=& 
(C^\mu+dx^\mu)(C^\nu+dx^\nu)
(\6_\mu A_\nu+\s0 12\, g_1t_{\mu\nu})
\nonumber\\
& &
-i\, (C^\mu+dx^\mu)(\lambda\sigma_\mu\5\xi-\xi\sigma_\mu\5\lambda)\ ,
\\
\7{\cal F} &=& (C^\mu+dx^\mu)(C^\nu+dx^\nu)\6_\mu a_\nu
-(C^\mu+dx^\mu)(S\sigma_\mu\5\xi+\xi\sigma_\mu\5S)
\eea
with $S$ and $\5S$ as in (\ref{S}).
These forms satisfy
\bea 
\4s\4c =\7{\cal F}\ ,\quad
\4s\4C={\cal B}-g_1\4Q\ ,\quad
\4s{\cal B}=g_1\4s\4Q={\cal H}
\label{app21}\eea
with $\4Q$ and ${\cal H}$ as in (\ref{app10}) and (\ref{app11}).
Using (\ref{app21}) and the technique of section 4 of \cite{sugra1},
one can check that the following total form is $\4s$-invariant:
\bea
\omega_{\cal A}&=&
\4c\, ({\cal B}^2-g_1\eta\lambda-g_1\5\eta\5\lambda-g_1\Xi D)
\nonumber\\
& &
+\Xi\, (\xi S\, \5\lambda \5\lambda-2\xi\lambda\, \5S \5\lambda
  +\5S\5\xi\, \lambda \lambda-2\5\lambda\5\xi\, S \lambda)
\eea
with $\Xi$ as in (\ref{Xi}). The volume form contained
in $\omega_{\cal A}$ is just the integrand of (\ref{an1}).
(\ref{an2}) is obtained by means of (\ref{eoma}) 
from the volume form contained in
\beq
\omega_{\cal A}+\4s\{\4c\, \4C({\cal B}+g_1\4Q)\}=
\7{\cal F}\4C{\cal B}+2g_1\4c\, \4C{\cal H}+\cdots\ .
\eeq

\end{document}